\documentclass[12pt]{article}

\usepackage[english]{babel}
\usepackage{float}

\usepackage[letterpaper,top=2cm,bottom=2cm,left=3cm,right=3cm,marginparwidth=1.75cm]{geometry}

\usepackage{graphicx}
\usepackage[colorlinks=true, allcolors=blue]{hyperref}
\usepackage{biblatex} 
\usepackage{url}

\usepackage{authblk}

\title{Music as an Asset Class} 

\author[1]{Sasha Stoikov}
\author[2]{Aadityaa Singla}
\author[1]{Umu Cetin}
\author[1]{Luis Alonso Cendra Villalobos}
\affil[1]{ORIE, Cornell University}
\affil[2]{Computer Science, Cornell University}

\usepackage{amsmath}               
  {
      \newtheorem{assumption}{Model}
  }
\begin{document}
\maketitle

\begin{abstract}

In the streaming era, music revenues distributed to rights holders have become more transparent. 
However, it is not yet clear how to quantify the risk and return characteristics of music royalty assets, as is done with equities. In this paper, we fit three discounted cashflow models to transactions on the Royalty Exchange platform. We use our best  model to backtest the one year and five year performance of music royalty assets, after transaction costs. We find that Life of Rights (LOR) music assets had risk and return characteristics comparable to stocks in the S\&P500, when held over 5 years. Since the performance of stocks and music assets are likely to be uncorrelated, this result may help investors assess this asset class within the context of a more traditional stock and bond portfolio. 

\end{abstract}

\section{Introduction}

The market for music royalties is illiquid: transactions are rare, there is significant information asymmetry and transaction costs are high. This poses significant challenges for pricing assets fairly and fitting models to past transactions. Without a model calibrated to transactions, it is impossible to estimate fluctuations in the value of an asset from one year to the other, using only song revenues data. 

As Shot Tower Capital, a firm that specializes in facilitating music transactions, states in their 2025 report \footnote{Shot Tower Capital. 2024 Annual Report. Baltimore: Shot Tower Capital, 2025.}, “The key valuation methodology for music catalogs and companies is a discounted cash flow analysis cross-checked against comparable market transactions". The first method, discounted cashflow analysis, takes into account past cash flows to estimate future cashflows. Though very popular and grounded in standard finance theory, this method can use a variety of potential factors (such as genre, age, vintage, or country) in a rather ad-hoc manner. The second method, comparing to recent market transactions, seems reasonable in theory. However it is challenging in practice since there are limited transaction data available for comparable artists.

Both these valuation methods, although useful for buyers and sellers to converge on a fair transaction price, don't say much about how good or bad an investment a music royalty asset is. Therefore, there is a pressing need for robust pricing models that can estimate the past returns of music royalties, even when market transactions are relatively sparse. 

In this paper, we fit three simple DCF models to 1295 transactions on the Royalty Exchange platform. With these calibrated models, we can use historical revenue data to estimate annual returns. We compute the expected return of an asset
$$r_i=\frac{p_{i+1}+c_i-p_i-t_i}{p_i}$$
where $p_i$ and $p_{i+1}$ are prices, $t_i$ is the transaction costs that the platform enacts (\$500 buyer fee and 8\% seller commission), and $c_i$ is the cashflow collected between period $i$ and $i+1$. The cashflows are usually observable on a quarterly basis, but  the prices $p_i$ are typically not observable, since transactions are rare.

 In section 2, we describe the Royalty Exchange data and highlight some of the stylized facts that differentiate music royalties from traditional assets like stocks and bonds. In section 3, we introduce three DCF models. We calibrate each model to actual transactions to fit the parameters. 
 In section 4, we use the best fitting model to backtest the risk/return characteristics of music royalties over a one year and five year period and compare them to equities over the same period. 
In section 5, we conclude with some limitations of our approach and suggestions for future work.

\begin{section}{Data}

Royalty Exchange is a transparent market for music royalty transactions. Before a rights holder sells on the platform, they must provide a history of past cashflows from a collection agency like ASCAP or BMI, and after some due diligence, buyers compete in a standard auction. After a primary sale happens, the asset may be traded in a secondary market with lower transaction costs. In our dataset, 1295 trades of which 1134 deals were primary sales were traded since 2017, for a face value of \$97M.

There are two main categories of data that Royalty Exchange publishes on the platform: deals and revenues. At the deal level, we know the price of the transaction, the LTM (last twelve months revenue), the LTY (last three years revenue, annualized), the age of the catalog and the deal terms (10Y, 30Y or LOR — Life of Rights). On the revenues side, we have a time series of quarterly revenues preceding each sale.

Since music royalty asset prices can vary significantly for different values of the LTM, market participants typically divide prices by the LTM to normalize transactions and express them in terms of 
a price-to-LTM ratio (also called a “multiplier” or "multiple"). 
In Figure \ref{fig:multipliers1}, we display the multipliers for 10Y and LOR as a function of LTM/LTY. 

\begin{figure}
    \centering
    \includegraphics[width=0.9\textwidth]{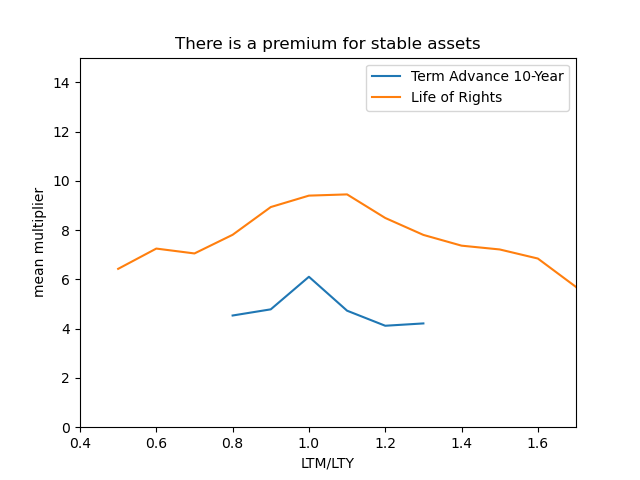}
    \caption{ Multipliers for 10Y and Life of Rights contracts as a function of LTM/LTY }
    \label{fig:multipliers1}
\end{figure}

For the LOR and 10Y contracts, we see that the market gives a premium to assets whose revenues are stable and the ratio of LTM/LTY is close to 1, compared to assets where recent revenues have been growing or shrinking. This suggests that in the context of a DCF model, there could be a risk-adjusted discount factor at play. In other words, the discount rate applied to cashflows that are more volatile is likely to be higher than for cashflows that are stable, where LTM and LTY are close to each other. 


In Figure \ref{fig:age}, we display the multipliers for 10Y and LOR as a function of the age of the catalog. Note that the multipliers are increasing in the age of a song.

\begin{figure}
    \centering
    \includegraphics[width=0.9\textwidth]{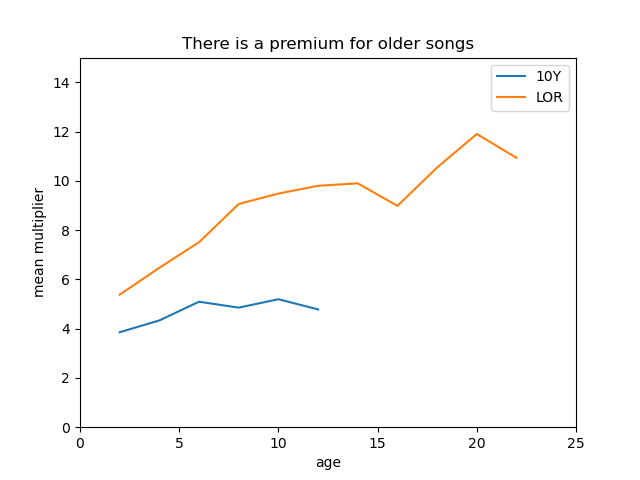}
    \caption{ Multipliers for 10Y and Life of Rights contracts as a function of song age }
    \label{fig:age}
\end{figure}


\end{section}

\begin{section}{Model}

To price an $n$-year royalty asset, we use a general discounted cashflow formula

\begin{equation}
    P^n=\sum_{i=1}^{n}   \frac{C}{(1+R)^i}
    \label{DCF}
\end{equation}
where $C$ is the expected yearly cashflow and $R$ is the discount rate. Different assumptions on $C$ and $R$ lead to different pricing models. In particular, we consider the following three models:
\begin{assumption}-\label{as:1}
  Expected music revenues will stay flat and the discount rate is constant. \begin{align*}
   C = LTM 
\end{align*}
\begin{align*}
   R = r 
\end{align*}
$$ M = \frac{P^n}{LTM}= \sum_{i=1}^{N} \frac{  1 }{(1+r )^i}$$

\end{assumption}

\begin{assumption}-\label{as:1}
  Discount rates are risk adjusted and cashflows are expected to stabilize at a level lower than the LTM. 
  \begin{align*}
   C = LTM *a
\end{align*}
   \begin{align*}
   R = r + k \times |LTM/LTY-1|
\end{align*}
$$ M = \frac{P^n}{LTM}= \sum_{i=1}^{N} \frac{  a }{(1+r+k\times |LTM/LTY-1|  )^i}$$
\end{assumption}

\begin{assumption}-\label{as:1}
  There is a premium for older songs
  \begin{align*}
   C = LTM *(a+b\times age)
\end{align*}
   \begin{align*}
   R = r + k \times |LTM/LTY-1|
\end{align*}
$$ M = \frac{P^n}{LTM}= \sum_{i=1}^{N} \frac{  a+b \times age }{(1+r+k\times |LTM/LTY-1|  )^i}$$
\end{assumption}

Note that Model 1 has a single parameter $\theta_1=(r)$, Model 2 has 3 parameters $\theta_2=(r,a,k)$ and Model 3 has 4 parameters $\theta_3=(r,a,k,b)$. Since each transaction has a traded multiplier, we may optimize over these parameters to obtain the least square error.

In other words, we run the following optimization
$$
\min_\theta \sum_{i=1}^{1295} (M_{traded}^i-M^i(\theta))^2
$$
and display the results in Table \ref{tab:3models}. 

\begin{table}[H]
\centering
\caption{Model Paramter Optimization}
\begin{tabular}{|l|l|l|l|l|l|}
\hline
\textbf{model} & \textbf{MSE} & \textbf{r} & \textbf{a} & \textbf{k} & \textbf{b}     \\ \hline
1              & 8.8          & 14.0\%     &        &        &   \\ \hline
2              & 7.7          & 7.6\%      & 0.69       & 0.071      &     \\ \hline
3              & 5.7          & 8.3\%      & 0.61       & 0.058      & 0.0098 \\ \hline
\end{tabular}
\label{tab:3models}
\end{table}

In the sequel, we use model 3, as it has a significantly lower mean squared error, compared to the simpler models. If we re-price all assets using model 3, we obtain Figure \ref{fig:multipliers_DCF_one}, which is qualitatively quite similar to Figure \ref{fig:multipliers1}. Furthermore, this model also leads to higher multipliers for older songs, see Figure \ref{fig:multipliers_DCF_two}, similar to the result in Figure \ref{fig:age}

\begin{figure} [H]
    \centering
    \includegraphics[width=0.9\textwidth]{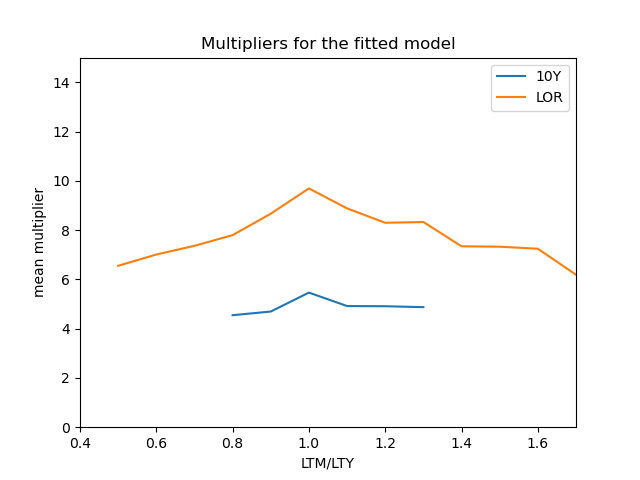}
    \caption{ Multipliers for 10Y and Life of Rights contracts as a function of LTM/LTY, for the model 3}
    \label{fig:multipliers_DCF_one}
\end{figure}

\begin{figure}[H]
    \centering
    \includegraphics[width=0.9\textwidth]{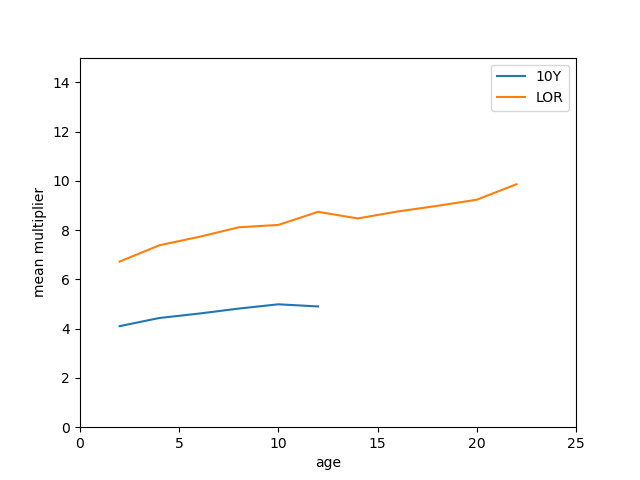}
    \caption{ Multipliers for 10Y and Life of Rights contracts as a function of age, for model 3}
    \label{fig:multipliers_DCF_two}
\end{figure}

\end{section}

\begin{section}{Backtesting}

How do the risk and return characteristics of a portfolio of songs compare to those of a portfolio of stocks? As mentioned in the introduction, transaction costs for songs on a platform like Royalty Exchange are around 8\%, much higher than the transaction costs of equities, which are typically expressed in basis points. Moreover, music assets have dividends that are often in the double digits, while stocks have negligible dividends. Finally, music assets tend to depreciate as revenues often decay, particularly for 10 year contracts that have a value of zero at expiry. However, in periods like 2017-2022, when the streaming grew steadily, Life of Rights assets have often appreciated, leading their investors to win on both counts, with high dividends and positive returns.




We now return to our formula in the introduction:
$$r_i=\frac{p_{i+1}+c_i-p_i-t_i}{p_i}$$
where $p_i$ and $p_{i+1}$ can now be computed using the model.
We decompose the returns into dividends:
$$d_i = \frac{c_i}{p_i}$$
 pure asset returns:
$$e_i = \frac{p_{i+1}-p_i}{p_i}$$
and transaction costs
$$f_i = \frac{t_i}{p_i}$$

Comparing the LOR and 10Y returns in Tables \ref{tab:LOR} and \ref{tab:10Y}, we find that the 10Y assets have high dividends and depreciate fast, while the LOR assets appreciate slightly, but offer lower dividends. This contrast highlights the fundamental tradeoff: 10Y contracts deliver yield up front but erode in value, while LOR contracts are longer-duration and more sensitive to discount rates, but can capture modest appreciation over time. 



The backtesting methodology simulates a one-year and 5 year buy-and-hold strategy. We assume an investor purchases an asset at time $t$ at the model-implied price $p_t$, calculated using only information available at that time (specifically, the LTM, the LTY and the age of the catalog). The investor then collects four quarters of actual realized cashflows $c_t$ and sells the asset at time $t+4$ at the model-implied price $p_{t+4}$, which is recalculated using the updated revenue information available at the end of the holding period. We incorporate transaction costs, $t_i$, in the form of a \$500 buyer fee (fixed) and an 8\% seller commission, which are deducted from returns to emulate the requirements of the Royalty Exchange platform. The five-year backtest follows a similar methodology to the one-year analysis: investors purchase at the model price in 2017, collect 20 quarters of actual cashflows, and sell at the model price in 2022.
 

\begin{table}[H]
\centering
\footnotesize
\caption{Life of Rights Assets}
\begin{tabular}{lccccccc}
\hline
\textbf{Metric (\%)} & \textbf{2017} & \textbf{2018} & \textbf{2019} & \textbf{2020} & \textbf{2021} &  \textbf{5-yr (total)}  & \textbf{5-yr (annualized)} \\ 
\hline
d - Median Dividends & 12.4\% & 11.3\% & 12.1\% & 11.8\% & 13.2\%  & 62.3\%  \\
e - Median Capital Gains & 10.7\% & 6.2\% & 6.1\% & 3.9\% & 8.4\%  & 25.4\% \\
f - Median TC & 8.9\% & 8.5\% & 8.5\% & 8.3\% & 8.7\%  & 10.0\% \\
\hline
r - Median Return & 12.5\% & 7.1\% & 9.0\% & 10.6\% & 12.9\%  & 82.64\% & 12.8\% \\
90th Percentile Return & 77.6\% & 72.4\% & 66.5\% & 47.9\% & 81.2\%  & 279.7\% & 30.6\% \\
10th Percentile Return & -23.1\% & -42.8\% & -28.3\% & -42.5\% & -25.5\%  & -7.84\% & -1.6\% \\
\hline
\end{tabular}
\label{tab:LOR}
\end{table}

\begin{table}[H]
\centering
\footnotesize
\caption{10-Year Assets}
\begin{tabular}{lccccccc}
\hline
\textbf{Metric (\%)} & \textbf{2017} & \textbf{2018} & \textbf{2019} & \textbf{2020} & \textbf{2021}  & \textbf{5-yr (total)}  & \textbf{5-yr (annualized)}  \\ 
\hline
d - Median Dividends  & 16.7\% & 17.3\% & 14.8\% & 13.1\% & 16.2\% & 90.3\% \\
e - Median Capital Gains & 3.3\% & -3.2\% & -1.6\% & -10.7\% & 6.4\%  & -40.5\% \\
f - Median TC & 8.3\% & 7.7\% & 7.9\% & 7.1\% & 8.5\% &  4.8\% \\
\hline
r - Median Return & 8.1\% & 2.0\% & 5.0\% & -7.1\% & 9.5\% &  42.5\% & 7.3\%  \\
90th Percentile Return & 64.2\% & 44.5\% & 42.1\% & 28.0\% & 51.1\% & 155.1\% & 20.6 \% \\
10th Percentile Return & -28.5\% & -27.5\% & -32.2\% & -37.7\% & -21.5\% &  -59.9\%  & -21.3\% \\
\hline
\end{tabular}
\label{tab:10Y}
\end{table}

\begin{table}[H]
\centering
\footnotesize
\caption{S\&P500 stocks}
\begin{tabular}{lccccccc}
\hline
\textbf{Metric (\%)} & \textbf{2017} & \textbf{2018} & \textbf{2019} & \textbf{2020} & \textbf{2021} &  \textbf{5-yr (total)}  & \textbf{5-yr (annualized)} \\ 
\hline
Median Return & 
22.5\% &  -3.3\% & 32.4\% & 13.2\% & 30.6\%   & 77.7 \% & 12.2\% \\
90th Percentile Return & 
58.1\% & 23.9\% &
64.5\% &
56.4\% &
 67.31\%  &  257.2 \% & 29.0\% \\
10th Percentile Return &

-3.2\% &
-30.3\% &
4.6\% &
-17.3\% &
0.2\%   & -15.3 \% & -3.3\% \\
\hline
\end{tabular}
\label{tab:SPY}
\end{table}

Both music asset types exhibit substantial spread between the top and bottom deciles, reflecting the unpredictable nature of music royalty performance. The return dispersion across the 10th and 90th percentiles is illustrated in the tables for LOR and 10Y assets. 

For life of rights contracts, the median annual return across 2017–2021 is 12.8\%, driven primarily by consistent dividend yields (11–13\% annually) and modest capital appreciation (typically between 4–11\% annually). This performance highlights that the majority of LOR returns come from recurring cashflows rather than speculative price movements. Ten year contracts exhibit lower median returns of roughly 7.3\% annually. These returns are primarily composed of higher dividend income (13–17\%) but are offset by frequent price depreciation as contracts approach expiration. This contrast highlights the structural differences in payout and resale profiles between perpetual and term-limited royalty assets. 

Finally, we compare these results with the median returns in the S\&P 500. To avoid survivorship bias, we selected the stocks in the index at the befinning of each backtest period and display the results in Table \ref{tab:SPY}. Note that the medians, 10th and 90th percentiles are comparable to the performance of LOR assets, over the course of a 5 year period, after transaction costs. However, from one year to the next, the median returns of LOR assets are much more stable, while the equity median returns change significantly from year to year. Indeed, one of the most often mentioned features of music assets is that they tend to be uncorrelated to financial markets. Our backtests are consistent with this idea.

\end{section}

\section{Conclusion}

Music royalties exhibit low correlation with equities, making them a potentially valuable addition to diversified portfolios. In this paper, we introduced three discounted cash flow models for valuing music royalty assets and demonstrated how they can be calibrated to observed transactions. Using these models, we backtested the performance of median music assets over one- and five-year horizons.


Our models, while useful, have important limitations. They do not explicitly incorporate characteristics such as genre, popularity,  volatility, or industry growth. We also assumed constant interest rates. Another concern is that our backtests are in-sample: model parameters were estimated using all transactions through 2025, whereas a real investor in, say, 2018 would not have had access to this information. 

Despite these caveats, our analysis suggests that Life of Rights assets outperform shorter-horizon 10-Year assets and, when held for five years, achieve risk-return profiles broadly comparable to equities. For music to mature as a true asset class, pricing models will need to move beyond financial history and incorporate measures of musical quality and cultural relevance. Ultimately, the drivers of over- or under-performance are not just financial, but artistic: how good the songs are, and how likely they are to expand their audience over time.

\printbibliography


\end{document}